\documentclass[aps,prl,superscriptaddress,twocolumn,floatfix]{revtex4}
\usepackage{amsmath,epsfig}
\usepackage{tabularx,multirow}

\begin{document}
\title{Theoretical and experimental analysis of H$_2$ binding in a prototype metal organic framework
material}
\author{Lingzhu Kong}
\affiliation{Department of Physics and Astronomy, Rutgers University, Piscataway, New Jersey 08854-8019}
\author{Valentino R. Cooper}
\affiliation{Department of Physics and Astronomy, Rutgers University, Piscataway, New Jersey 08854-8019}
\affiliation{Materials Science and Technology Division, Oak Ridge National Laboratory, Oak Ridge, Tennessee 37831-6114}
\thanks{Permanent address for V.R.C. since 9/16/08.}
\author{Nour Nijem}
\affiliation{Department of Materials Science and Engineering, University of Texas at Dallas, Richardson, Texas 75080}
\author{Kunhao Li}
\affiliation{Department of Chemistry and Chemical Biology, Rutgers University, Piscataway, New Jersey 08854-8087}
\author{Jing Li}
\affiliation{Department of Chemistry and Chemical Biology, Rutgers University, Piscataway, New Jersey 08854-8087}
\author{Yves J. Chabal}
\affiliation{Department of Materials Science and Engineering, University of Texas at Dallas, Richardson, Texas 75080}
\author{David C. Langreth}
\affiliation{Department of Physics and Astronomy, Rutgers University, Piscataway, New Jersey 08854-8019}

\begin{abstract}
Hydrogen 
adsorption by  the metal organic framework (MOF) structure Zn$_2$(BDC)$_2$(TED)
 is investigated using a combination of experimental and theoretical
 methods.
By use of the nonempirical van der Waals density-functional (vdW-DF) approach, 
it is found that the locus of deepest H$_2$ binding positions lies within
two types of narrow channel.  The energies of the most stable
binding sites, as well as the number of such binding sites,
are consistent with the values obtained from  experimental
adsorption isotherms and heat of adsorption data.
 Calculations  of the  shift of the H--H stretch
frequency when adsorbed in the MOF give a value of approximately
$-30$ cm$^{-1}$ at the strongest binding point in each of the
two channels. Ambient temperature infrared
absorption spectroscopy measurements give a hydrogen  peak
centered at 4120 cm$^{-1}$, implying a shift consistent with the theoretical 
calculations.
\end{abstract}

\pacs{81.05Zx, 84.60.Ve}
\keywords{hydrogen storage; energy storage; metal organic framework; MOF; vdW-DF}
\maketitle

Hydrogen storage technology is one bottleneck for the utilization of 
hydrogen as an energy source for mobile applications. 
Metal-organic frameworks (MOFs) comprise a rather new class of porous materials
in which metal ions or clusters are linked by organic units~\cite{Eddaoudi2001}.
The scaffold structure and large apparent surface
area make these materials potential candidates for hydrogen storage 
applications.  
One such material, MOF-5, was 
shown to adsorb 1.3 wt.-\% of hydrogen at 77 K and 1 atm~\cite{Rowsell2004}.
A recent effort with Zn$_2$(BDC)$_2$(TED) reported an
uptake of 2.1 wt.-\%~\cite{Lee2007}. Here BDC is benzenedicarboxylate (C$_8$H$_4$O$_4$)
and TED is triethylenediamine (C$_6$H$_{12}$N$_2$).
To rationally design or modify MOF  structures to meet
storage needs, it is critical to understand the interaction
between the MOF matrix and the adsorbed hydrogen molecules.
Furthermore, the quest to understand the physisorption and possible dissociative chemisorption
of H$_2$ within a MOF is a fascinating problem in its own right.  In the present
letter, we show how  a relatively recently developed nonempirical theoretical method
\cite{Dion2004,Thonhauser2007} and infrared absorption 
measurements \cite{Chabal1984} can be combined with 
isotherm and heat of adsorption data \cite{Lee2007} 
to obtain a detailed picture of the adsorption of H$_2$ in MOFs.
For this purpose we use Zn$_2$(BDC)$_2$(TED) \cite{Lee2007} as a prototype.

Spectroscopic and theoretical studies have confirmed that H$_2$ is weakly 
bound in MOFs and the binding is mainly from  long-range dispersive 
interactions~\cite{Centrone2005}. The nonlocal electron correlation
in this type of interaction makes accurate and efficient electronic
structure calculations very  difficult. Quantum chemical methods that
account for these interactions typically scale poorly with system size
and are limited to only fragments of the true MOF structure.
On the other hand, ordinary density functional theory (DFT) methods, which are efficient
and scale well with system size, 
fail to reproduce
the correct behavior of the van der Waals interactions, which are important
in these systems~\cite{Heine2004}. For example, an H$_2$ binding energy 
of 21.7 meV was obtained for the main H$_2$ adsorption sites in MOF-5 
by use of the generalized gradient approximation (GGA) within DFT~\cite{Mueller2005},
while the experimental value is 49--54 meV~\cite{Long2005}.

An alternative, a van der Waals density functional (vdW-DF), has been 
developed by our group and collaborators~\cite{Dion2004,Thonhauser2007}. It incorporates
the van der Waals interaction into a fully nonlocal and nonempirical density functional
for the correlation energy,
which retains  ordinary DFT's good description of covalent bonding.
The method has been successfully applied to various sparse 
systems~\cite{Dion2004,Thonhauser2007,Several_key_vdW-DF_papers}. Here we begin by benchmarking
vdW-DF  on
the binding of H$_2$ to an isolated  BDC linker, where there exist
 calculations via well established quantum chemical methods \cite{QCcalcs,sugara}.
It is found that H$_2$ binds most strongly when pointed along the  symmetrically placed  axis
through the center of the benzene ring, and sitting a distance a little over 3 {\AA} away
with a binding energy of approximately 4 kJ/mol. We find that while 
a standard GGA calculation using the PBE functional \cite{PBE1996} fails to predict
more than a third of the interaction energy, the vdW-DF \cite{tech_detail} 
not only
gives the correct binding energy, but  quite well reproduces the whole binding curve \cite{sugara}
 and the potential energy landscape \cite{sugara}, though, as is typical 
 \cite{Dion2004,Thonhauser2007,Several_key_vdW-DF_papers}, the intermonomer equilibrium
 distance is slightly too large.
In the full MOF, we  find that this site's binding energy increases
by 3 kJ/mol  due to long range interactions with other
parts of the MOF---a very nonlocal effect that cannot  be
reproduced by standard density functionals.
 More spectacularly this binding point of the fragment becomes
only a saddle point in the potential energy surface of the MOF, 
showing that calculations on fragments
alone  do not necessarily give relevant information about the binding sites
in the full MOF, or even useful information as to where to look for them.

\begin{figure}[ht]
\centerline{\epsfig{file=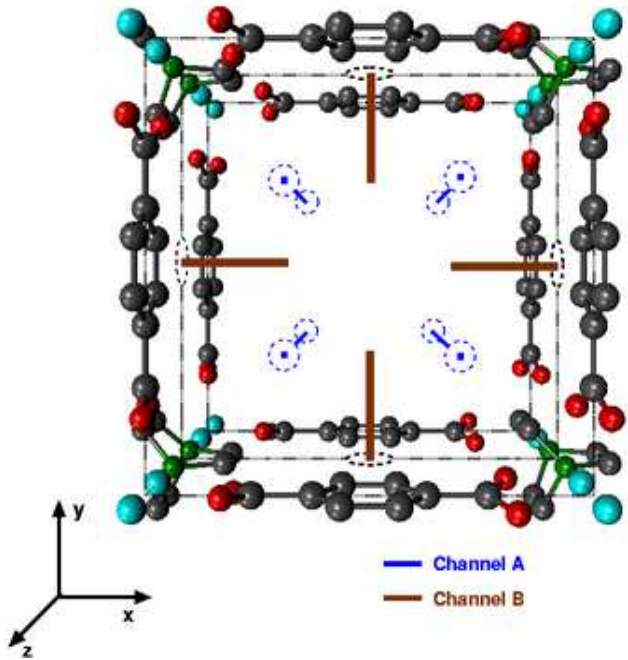,width=0.47\textwidth}}
\caption{A perspective top view of the unit cell of the MOF. The BDC linkers
and the metal atoms are evident in this view, while the TED pillars are
more clearly seen in the two subsequent figures. Atomic color scheme:
black C, green N, red O, blue-green Zn; H atoms are not shown.
The locations of Channels A (thin blue) and B (thick brown) are schematically
indicated. The dashed circles indicate places where the
respective channels continue into the adjoining unit cell.\label{fig:MOF1}}
\end{figure}

\begin{figure}[ht]
\centerline{\epsfig{file=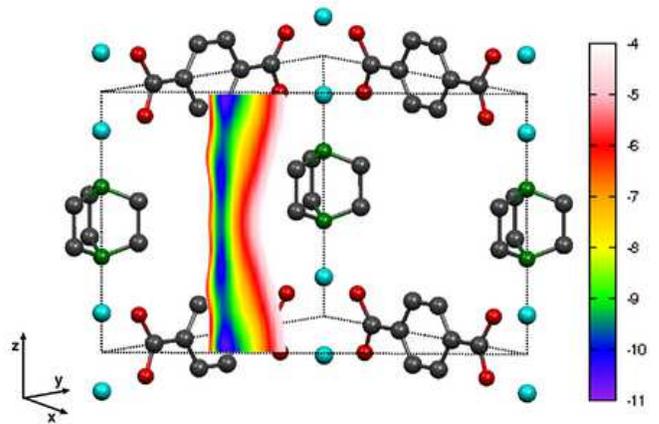,width=0.49\textwidth}}
\caption{A diagonal side view of half  the unit cell, showing a contour energy map
of one of the four instances of Channel A.  The H$_2$ molecule was aligned in the $z$ direction.
The units of the color scale
are kJ/mol. \label{fig:MOF2}}
\end{figure}

\begin{figure}[b]
\vspace*{-11pt}
\centerline{\epsfig{file=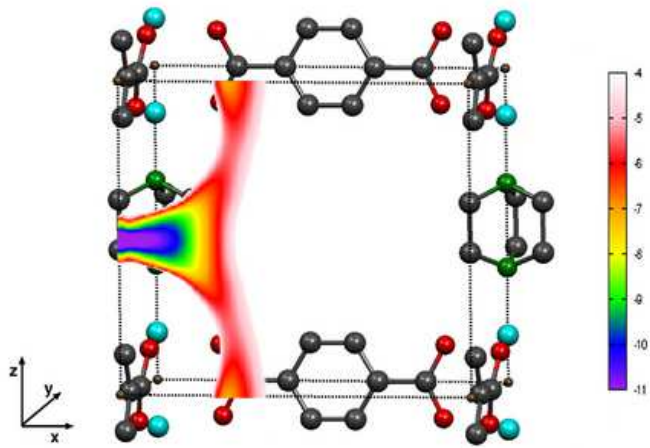,width=0.49\textwidth}}
\caption{A side view of half the unit cell, showing one of the four instances
of Channel B, each of which extend an equal amount into the respective
neighboring unit cell. The contour map color scale and the H$_2$
orientation are the same as
for  Fig.~\ref{fig:MOF2}.\label{fig:MOF3}}
\end{figure}

The basic MOF structure was taken from experiment \cite{Lee2007}. 
Based on the single crystal data of the guest-free structure \cite{kim2004}, the C in
TED and O in BDC have a 4-fold and 2-fold disorder, respectively. A simplified structure
eliminating the disorder is used in the actual calculations \cite{kim2004}, as shown
in Fig.~\ref{fig:MOF1}. The underlying Bravais lattice is tetragonal with
$a=b=10.93$ {\AA} and $c=9.61$ {\AA}, with a basis consisting of one
formula unit of Zn$_2$(BDC)$_2$(TED).
Our vdW-DF calculations found that the strong H$_2$ binding regions are
located in two types of channel (see Fig.~\ref{fig:MOF1}). One such channel type runs in the $c$ direction
through the length of the crystal, with four such channels entering and leaving
each unit cell. We denote these as Channels A. An energy contour map of one of them is
shown in Fig.~\ref{fig:MOF2}.
These
four channels are not quite equivalent in the calculated model because of
the differing faces presented by the nearest TED and oxygen twist, but the differences
are relatively small ($\sim$1~kJ/mol).
A second type of channel, which we refer to as Channel B, runs in the $a$ or $b$ direction.
These run only from one unit cell (as we draw it here) to the next, because they
would otherwise run through the center of the cell, which is a weak binding region.
One of the halves of Channel B is illustrated in Fig.~\ref{fig:MOF3}. There are
four halves of Channel B per unit cell, i.e.\ two Channel B's per cell. What is
clear from the the maps in the Figs.~\ref{fig:MOF2} and \ref{fig:MOF3}, is that the mean
binding energy in the maximum regions is 10 kJ/mol in round numbers for each of the
channels, with Channel B perhaps a little higher. Although the maximum binding
energies in Channels B are affected only very slightly ($\sim$0.2 kJ/mol) by the
orientation of the  nearest TED and oxygen components, the widths of these channels in
the $xy$ plane (not illustrated) are more strongly affected.
It is interesting
that unlike what is apparently true for some isoreticular MOFs (IRMOFs) \cite{Rowsell2005},
there is little binding specifically associated with the metal atoms or in their vicinity.

 There are two caveats with respect to the predicted  binding of $\sim$10 kJ/mol.
  First, the binding is dependent
on the H$_2$ orientation (see Table \ref{table:frequency}), with a variation of $\sim$1 kJ/mol. Second,
the channels are narrow, and H$_2$ is a quantum mechanical object. Its ground
state energy will be higher than the potential minimum. A full treatment is
beyond the scope of this letter, but harmonic zero point energy estimates
indicate that this effect could reduce the effective binding by
1 to 2 kJ/mol per H$_2$. 

The above results are consistent with the experimental H$_2$ uptake
curves at $T=77$ K and 87 K \cite{Lee2007},
which we found to be Langmuir isotherms to a high degree of accuracy.
The Langmuir isotherm is simply a plot of a Fermi function against
$\exp (\mu /kT)$  ($\mu$ is the chemical potential),
combined with the realization that $\exp (\mu /kT)$
is proportional to the pressure $P$ outside the MOF. The fact that a single
Langmuir isotherm (rather than a linear combination
of them) fits the uptake data implies that the important adsorption sites all have
about the same binding energy, just as we find. The experimental
binding energy can be determined both from the isotherm fit
and independent heat of adsorption measurements \cite{Lee2007}: each give
a value of $\sim$\relax$7$ kJ/mol. When the quantum corrections
mentioned above as well as thermal excitations involving them
are accurately calculated and applied, we expect that
our well-bottom value of $\sim$\relax$10$ kJ/mol will be significantly
reduced and closer to experiment.  Finally, our fit to the
Langmuir isotherms gives $\sim$13 adsorption sites per unit cell.
An inspection of Figs.~\ref{fig:MOF2} and \ref{fig:MOF3} implies a total of 12,
provided that we assign two per unit cell to each of the 4
Channels A per cell, and 1 to each of the Channel B halves.
More importantly, the magnitude of the interaction
energy increases approximately linearly as we load hydrogen molecules on
these sites one by one.

We now turn to the calculation of the H$_2$ stretch frequencies. We carried
out a series of calculations varying the bond length of H$_2$,
with the center of H$_2$ and the host
atoms fixed at their equilibrium positions. The  resulting
 potential energy curve for each configuration
 was used in the Schr\"{o}dinger equation
to obtain the eigenvalues and excitation frequencies~\cite{Cooley1961}.
At each
of the two minimum positions, we performed the calculations for all three
molecular orientations along $x$, $y$, and $z$. For the Channel A
minimum, we calculated the stretching frequency along the 
body-center 
diagonal
as well. 
We express our results in terms  of the shifts from our
calculated value of the frequency for free H$_2$ of 4160 cm$^{-1}$.
Although the latter is accidentally more accurate than expected for
the level of approximation used, no use is made of this fact, and
we consider only the frequency shift upon adsorption to
be relevant for the comparison with experiment.

The vibrational frequency of the hydrogen molecule shifts downward when
absorbed into the MOF matrix, as shown in Table~\ref {table:frequency}.
The relative change from the free H$_2$ value, 
 is mainly due to the van der Waals interaction, which is
 weak.  The shifts predicted by an application of the GGA were found
 to be substantially smaller still. The shifts that we predicted with
 the vdW functional were
$-28$ cm$^{-1}$ and $-25$ cm$^{-1}$ for the preferable binding
orientations in Channels A and B respectively and are smaller
in magnitude for
other molecular orientations. The interaction
energies for different orientations are within $\sim$1~kJ/mol
for each channel, which indicates that all
the orientations will be well populated at room temperature.
Furthermore, the translational states, which have a 
range of frequencies of up to $\sim$100 cm$^{-1}$,
will be well occupied, implying 
gain and loss contributions of comparable amplitudes.
As a result, we expect the frequency peak to be substantially
broadened at such ambient temperatures.

\begin{table}[h]
\caption{ Calculated vibrational frequency shifts and interaction energies
          for various orientations at the two types of binding site. The largest binding
	  Channel A sites are at the intersections of that channel with the cell body
	  diagonals (see Fig.~\ref{fig:MOF2}), while those for Channel B are at the
	  cell boundaries (see Fig.~\ref{fig:MOF3}).
	  The frequency shifts are relative to our calculated values for free H$_2$ of
	  4160 cm$^{-1}$. }
\label{table:frequency}
\begin{ruledtabular}
\begin{tabular*}{1.0\textwidth}{@{\extracolsep{\fill}}ccccc}

 \multirow{2}{*}{}   & H$_2$ & { $ \omega $ shift}  & Energy \\
 &orientation&cm$^{-1}$& (kJ/mol)\\

\hline
  \multirow{4}{*}{Channel A}
     &  diagonal  &  $-28$ &  $-10.1$ \\
     &  x         &  $-17$ &  $ -9.1$ \\
     &  y         &  $-21$ &  $ -9.5$ \\
     &  z         &  $-15$ &  $ -8.7$ \\

\hline
 \multirow{3}{*}{Channel B}
 & x              &$-25$ & $-10.9$  \\
 & y              &$-15$ & $ -9.6$  \\
 & z              & $-4$ & $ -9.7$  \\

\end{tabular*}
\end{ruledtabular}
\end{table}

\newcommand{\HH}{H$_2$}

Infrared absorption spectroscopy is particularly useful for studying the incorporation 
of {\HH} molecules in semiconductors, as was demonstrated for amorphous 
silicon \cite{Chabal1984}. The onset of IR activity and the position of 
the {\HH} internal vibrational mode are both sensitive measures 
of {\HH} interaction with the matrix. The actual measurements are difficult 
because the absorption cross section of {\HH} is expected to be so weak that 
overtones and combination bands of the MOF material will interfere with the measurement. 

\begin{figure}[h]
\begin{center}
\epsfig{file=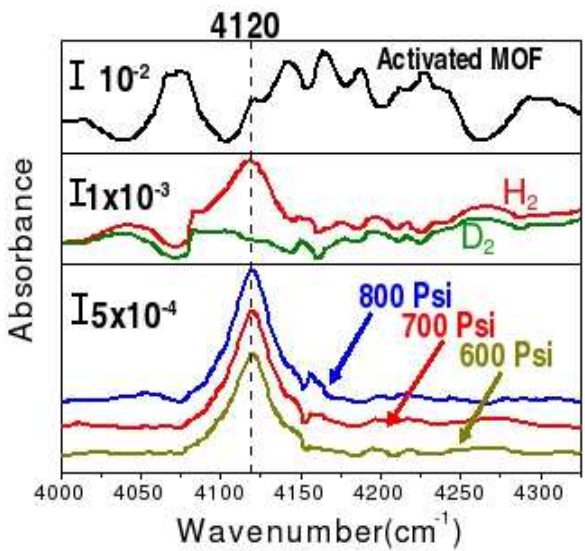,width=0.49\textwidth}
\caption{Infrared absorbance of the Zn$_2$(BDC)$_2$(TED) MOF at room temperature. 
Top panel: clean MOF at ambient pressure; Middle panel: MOF at 800 psi of H$_2$
(upper curve) or D$_2$ (lower curve); Bottom panel: difference between the
H$_2$ and D$_2$ spectra at several pressures.
 \label{fig:IR1}}
\end{center}
\end{figure}

For these experiments,  Zn$_2$(BDC)$_2$(TED) powder samples were prepared solvothermally \cite{Lee2007}.
A small amount (10 mg) was lightly 
pressed on a KBr pellet, the constitutive solvent removed by activation (annealing for 10 h under 
10$^{-2}$ Torr vacuum) and the spectrum was recorded (using a clean KBr pellet as reference). The top panel of Fig.~\ref{fig:IR1} shows the spectrum associated with overtones and combination bands of the clean MOF. Upon loading 
with high pressure He gas, these bands are perturbed, as evidenced by spectral shifts, due to minor rearrangement of 
the organic ligands. Consequently, in order to obtain a clean spectrum of {\HH} molecular absorption, D$_2$ molecules 
were used as a background for the {\HH} absorption experiments, as shown in the middle spectra in Fig.~\ref{fig:IR1}. 
The difference obtained is shown in the bottom three spectra for selected pressures (600, 700 and 800 psi). 

\newcommand{\icm}{cm$^{-1}$}

The main observation is the presence of a well-defined band centered at 4120 {\icm}, 
which is composed of 3/4 \mbox{ortho} and 1/4  \mbox{para} hydrogen (room temperature distribution), 
hence the asymmetry towards higher frequencies. The measured shift of $-35$ {\icm} 
from the unperturbed ortho (4155 {\icm}) 
and para (4161 {\icm}) frequencies is in good agreement with the calculated $-28/25$
{\icm} shifts. Furthermore, the integrated areas 
(measured from 300 to 800 psi)
are linear with pressure,
which is expected since, even at 800 psi, 
the average H$_2$ loading per unit
cell is  
only $\sim$1 {\HH} out of $\sim$12 possible sites, well within the 
linear regime for sequential loading found by the calculations. 
The observed vibrational band is $\sim$54 {\icm} broad, consistent with 
phonon and thermal broadening at room temperature, and with tails extending 
over 50 {\icm}, which may arise from the translational states.

\vspace*{-11pt}
To summarize, we have shown that a combination of experimental
and theoretical methods gives a consistent and  accurate
picture of {\HH} binding in a prototype MOF structure. This
picture will become still more precise when both low temperature
infrared absorption can be performed, and  the calculations of 
quantum effects on the motion of {\HH} within the MOF are complete.
The methods are now ripe for application to larger MOF structures
with larger {\HH} binding energies and hopefully will give 
additional clues about structural and chemical changes that can increase
these energies even further.

Work supported by DOE-DE-FG02-08ER46491.  
Work by V.R.C. supported by  NSF-DMR-0456937 at Rutgers until 9/15/08 and by
DOE, Division of Materials Sciences and Engineering at ORNL after 9/15/08.

\end{document}